\documentstyle[12pt,epsf]{article}
\begin{document}
\baselineskip=24pt

\title{
\vspace{-2.0cm}
\begin{flushright}
{\normalsize UTHEP-313}\\
\vspace{-0.3cm}
{\normalsize August 1995}\\
\end{flushright}
\vspace*{3.0cm}
{\Large Finite-Temperature Phase Structure of Lattice QCD\\
        with Wilson Quark Action\vspace*{0.5cm}}
}
\author{S. Aoki, A. Ukawa and T. Umemura\\ \\
{\it Institute of Physics, University of Tsukuba,}\\
{\it      Tsukuba, Ibaraki-305, Japan}
}
\date{}
\maketitle







\vspace*{1cm}
\begin{abstract}
\baselineskip=24pt
The long-standing issue of the nature of the critical line of lattice
QCD with the Wilson quark action at finite-temperatures, defined to be the line
of vanishing pion screening mass, and its relation to the line of
finite-temperature chiral transition is examined.  Analytical and numerical
evidence are presented that the critical line forms a cusp at a finite
gauge coupling, and that the line of chiral transition runs past the tip of
the cusp without touching the critical line.
Implications on the continuum limit and the flavor dependence of chiral
transition are discussed.
\end{abstract}



\newpage

Elucidating the nature of chiral transition separating the high-temperature
quark-gluon plasma phase from the low-temperature hadron phase has
been one of the focal points of effort in recent numerical
simulations of full lattice QCD including light dynamical quarks.  Much of work
in this direction has employed the Kogut-Susskind quark
action since it retains a $U(1)$ subgroup of chiral symmetry.  On the other
hand, studies with the Wilson quark action are less well
developed in spite of the effort over the years\cite{earlywork,qcdpaxtwo,milc}.
In fact, two issues, which are fundamental for understanding the
chiral transition with this action, have not been fully clarified to
date.   Both of the issues originate from the difficulty of identifying the
chiral limit of massless quark due to the presence of explicit chiral symmetry
breaking in the action,
which is introduced to avoid the well-known species doubling problem.

For the case of zero temperature pion mass has been conventionally employed to
deal with this problem.  Let $\beta=6/g^2$ be the inverse gauge coupling and
$K$ the hopping parameter which controls quark mass.  There is much evidence
from
analytical considerations and numerical simulations of
hadron masses that the pion mass vanishes along a line $K=K_c(\beta)$, the
critical line, which runs from $K_c(\beta=0)=1/4$ in the strong-coupling limit
to $K_c(\beta=\infty)=1/8$ in the weak-coupling limit.
At finite temperatures, one is then naturally led to define the
critical line in terms of the pion screening mass extracted from the
exponential decay of pion propagator for a large spatial separation.
The first issue in understanding the chiral transition is how the
critical line thus defined behaves as a function of $\beta$ and the
temporal lattice size $N_t$, and how it is related to the critical line at
zero temperature.

The second issue concerns the relation of the critical line and
the line of finite-temperature chiral transition $K=K_t(\beta)$, the thermal
line.  Naively one would expect the thermal line to meet the critical line at
some value $\beta=\beta_c$.  In this
case, the region bounded by the thermal line and the critical line for
$\beta\leq\beta_c$ represents the low-temperature phase and the other region
the
high-temperature phase. However, initial simulations\cite{earlywork} failed to
find clear indications of such a behavior; the two lines appeared to run
almost parallel toward the region of strong coupling, down to $\beta=5.0-4.5$
for the case of $N_f=2$ flavors.  These results raised the question if the two
lines meet at all, which has led to subsequent studies\cite{qcdpaxtwo,milc}.
In particular the QCDPAX Collaboration carried out simulations down to the
strong
coupling limit $\beta=0$ and concluded that
the two lines meet at $\beta\approx 3.9-4.0$\cite{qcdpaxtwo}.
However, their result has not satisfactorily answered the question in our view
since the nature of the critical line at finite temperatures we have addressed
above has not been clarified in their work.

In this article we report results of our study on the two fundamental
issues summarized above.  Our analyses are based on the idea of spontaneously
breakdown of parity and flavor as a characterization of the critical line,
which
has been put forward by one of us\cite{aoki}.  The phase structure we found
exhibits some unexpected features which we substantiate with analytical
arguments and hybrid Monte Carlo simulations of $N_f=2$ flavor full QCD with
the
Wilson quark action on an $8^3\times 4$ lattice.  We shall also discuss
implications of
our results on the continuum limit of the $N_f=2$ chiral transition and
extensions for the case of $N_f\geq 3$.

\begin{figure}[h]
\centerline{\epsfxsize=14cm \epsfbox{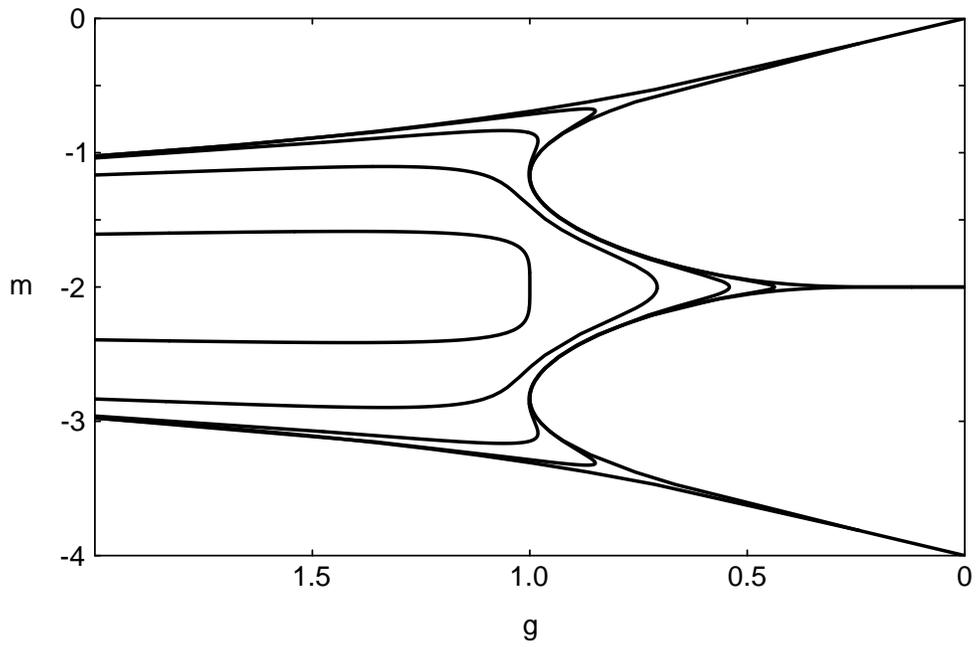}}
\caption{Critical line for the lattice Gross-Neveu model on $(g,m)$ plane.
Temporal lattice size equals $N_t=2, 4, 8, 16$ and $\infty$ from inside
to outside.}  \label{fig:fig1} \end{figure}

The starting point of our analysis is the Gross-Neveu model in two dimensions
formulated with the Wilson quark action\cite{eguchinakayama}.
Except for confinement, this model has
the features quite similar to those of QCD: asymptotic freedom, spontaneous
breakdown of chiral symmetry and its restoration at a finite
temperature.  In the large $N$ limit, the pion mass in this model is
analytically
calculable in the saddle-point approximation.
In Fig.~\ref{fig:fig1} we plot the critical line corresponding to $m_\pi=0$ on
the $(g,m)$ plane for the temporal lattice sizes $N_t=2, 4, 8, 16$ and
$\infty$ where $g$ and $m$ are the bare coupling and quark mass, the latter
related to the hopping parameter through $K=1/(2m+4)$.
The result for $N_t=\infty$, which has been known for some
time\cite{aoki}, shows that the critical line forms three cusps which
touch the weak-coupling limit $g=0$ at $m=0, -2$ and $-4$.  The region
outside of the critical line is a normal phase with $m_\pi\ne 0$, while the
inside is the phase with spontaneous breakdown of parity characterized by
$<\bar\psi\gamma_5\psi >\ne 0$.  Conventionally one constructs the continuum
limit of the model close to the point $(g,m)=(0,0)$ along the
branch of the critical line extending over $\infty\geq g\geq 0$.  The existence
of additional branches converging toward $(g,m)=(0,-2)$ and $(0,-4)$ arises
from a rearrangement of the doubler and massless fermion spectrum
as $m$ is varied.

For understanding the phase structure at finite
temperatures, the crucial feature revealed in Fig.~\ref{fig:fig1} is that the
three cusps of the critical line retract from the weak-coupling limit $g=0$
for finite temporal lattice sizes, forming a single continuous line which
shifts
toward strong coupling as $N_t$ becomes smaller.  Thus, for a finite $N_t$, the
critical line is absent for sufficiently weak coupling.  Also noteworthy is the
fact that, even in the range of $g$ where the critical line exists, the
position of the critical line depends on $N_t$, albeit only slightly
for large values of $N_t$.

The close resemblance of the Gross-Neveu model and QCD suggests that a
similar behavior of critical line holds for QCD except that the number of cusps
should increase to five because of the difference in the number of dimensions.
For the zero temperature case, the available evidence supporting this
possibility
is as follows: (i) In the
strong-coupling limit,  analytical results in the large $N$ limit  for chiral
observables such as $<\bar\psi\gamma_5\tau_3\psi >$ and associated
susceptibilities
show spontaneous breakdown of parity and flavor for $\vert
K\vert\geq 1/4$\cite{aoki}.  Results of a numerical simulation for the $N_f=2$
full
QCD system show good agreement with the analytic predictions\cite{auu}.  (ii) A
quenched calculation of the number of conjugate gradient
iterations needed for inverting the Wilson quark matrix over the range
$\beta=5.0-6.0$
revealed the presence of three peaks in the region $K\geq 0$ for $\beta\geq
5.6$\cite{auu}.
(iii) A measurement of hadron masses at $\beta=6.0$ in
quenched QCD on a $10^3\times 20$ lattice found clear evidence for three more
values of $K_c$ beyond the
conventional one at which pion mass vanishes\cite{auu}.  The parity broken
phase situated between two successive critical values is narrow, being of order
$\delta K_c\approx 0.05-0.1$.

For the finite temperature case, an apparent disappearance of
the critical line toward weak coupling
has been noticed in previous studies.  In the $N_f=2$ results of the
MILC Collaboration\cite{milc} on an $N_t=4$ lattice in the range
$\beta=4.9-5.3$, the pion screening mass initially decreases toward larger
values of $K$, but it increases beyond the thermal line $K_t(\beta)$.
A similar behavior is
seen in the data of the QCDPAX Collaboration at
$\beta=4.5-4.3$\cite{qcdpaxtwo}.
On the other hand,
the conventional critical line has been shown to remain at $\beta=0$ and
$3.5$\cite{qcdpaxtwo}.
Therefore, confirming a structure similar
to that of Fig.~\ref{fig:fig1} for QCD
requires evidence that the conventional critical line turns back toward strong
coupling at some value of $\beta$.

In order to examine this point, we have carried out hybrid Monte
Carlo simulations of $N_f=2$ full QCD with the Wilson quark action on an
$8^3\times 4$ lattice.  Runs were made in the range $3.0 \leq\beta\leq 5.3$ and
$0.15\leq K\leq 0.31$, which covers the region beyond the
conventional critical line, typically in steps of $\delta\beta=0.5$ and
$\delta K=0.01-0.02$.  The conjugate gradient method was
employed for inverting the Wilson quark matrix with the stopping
condition of $\vert\vert\xi-D^\dagger Dx\vert\vert^2/N<10^{-5}$ with
$N=\vert\vert x\vert\vert^2$ or $N=12V$ with $V$ the lattice volume. In order
to
maintain the acceptance at the level of $70-80$\% or more, the
hybrid Monte Carlo step size was decreased from $\delta\tau=0.02$ to
$0.00125$ as $\beta$ is decreased or $K$ is moved closer to the critical
line. Choosing the interval of $0.5$ or $0.25$ to be the unit trajectory, at
least
$50-100$ trajectories were generated after thermalization with
local observables measured for each trajectory. At $\beta=4.0$ and $3.5$ hadron
propagators were evaluated at every two trajectories by periodically doubling
the
lattice in spatial directions  $8^3\times 4\to (8\times 2)\times
8^2\times 4$.

\begin{figure}[h]
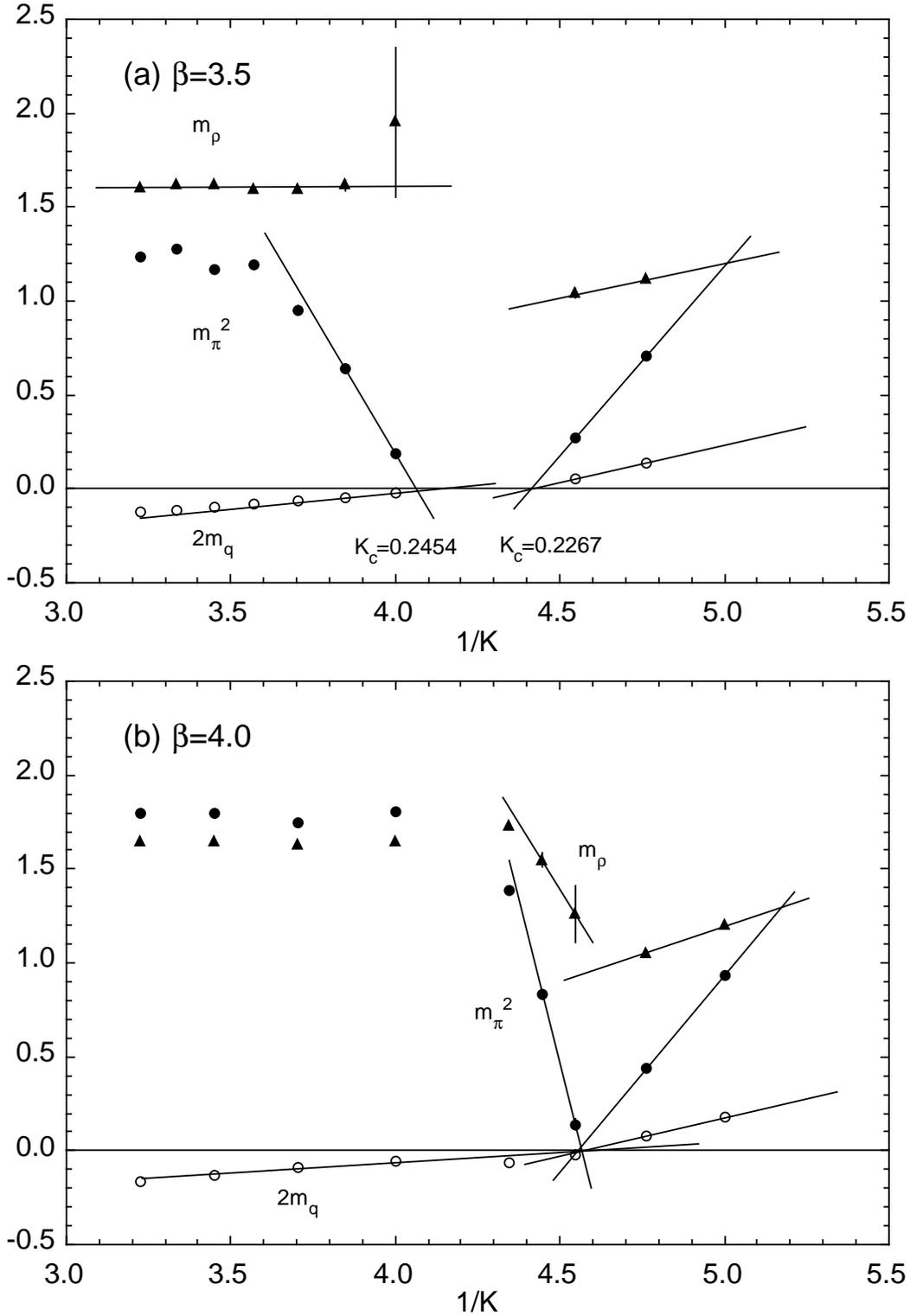

\centerline{\epsfxsize=14cm \epsfbox{fig2a.epsf}}
\centerline{\epsfxsize=14cm \epsfbox{fig2b.epsf}}
\caption{$\pi$ and $\rho$ screening masses and quark mass for $N_f=2$ full QCD
with Wilson quark action obtained on an $8^3\times 4$ lattice periodically
doubled in one of spatial directions.}
\label{fig:fig2} \end{figure}

In Fig.~\ref{fig:fig2}(a) we present our result for $\pi$ and $\rho$ screening
masses at $\beta=3.5$ as a function of $1/K$.  Also plotted is the quark mass
defined via the axial vector Ward identity\cite{ward}.
To the right of the figure is the low-temperature phase bounded by the
conventional critical line at $K_c\approx 0.2267$ where the pion mass squared
and  the quark mass linearly vanish almost simultaneously.
Clearly there exists another critical line at
$K_c\approx 0.2454$ below which the pion and quark masses behave in a reverse
manner.
Fig.~\ref{fig:fig2}(b) shows how the behavior changes at $\beta=4.0$.
We observe that the gap between the two critical values has either become
extremely narrow or disappeared.  These results lead us to believe that we have
identified one of the cusps of the critical line, with the turning point
located in the vicinity of $\beta=4.0$.

Let us now discuss the question whether the thermal line
$K_t(\beta)$ crosses the critical line $K_c(\beta)$.  In the light of our
results above, it should be clear that this can not happen with our definition
of
$K_c(\beta)$; the part of the critical
line beyond the crossing point, assuming it exists, should be in the high
temperature phase, where we expect the pion screening mass to be finite, while
it vanishes along the entire critical line by definition.
In other words the region close
to the critical line has to be in the low-temperature phase even after it turns
back
toward strong coupling.  This means that the thermal line should run past the
turning
point of the critical line and continue toward larger values of $K$.

\begin{figure}[h]
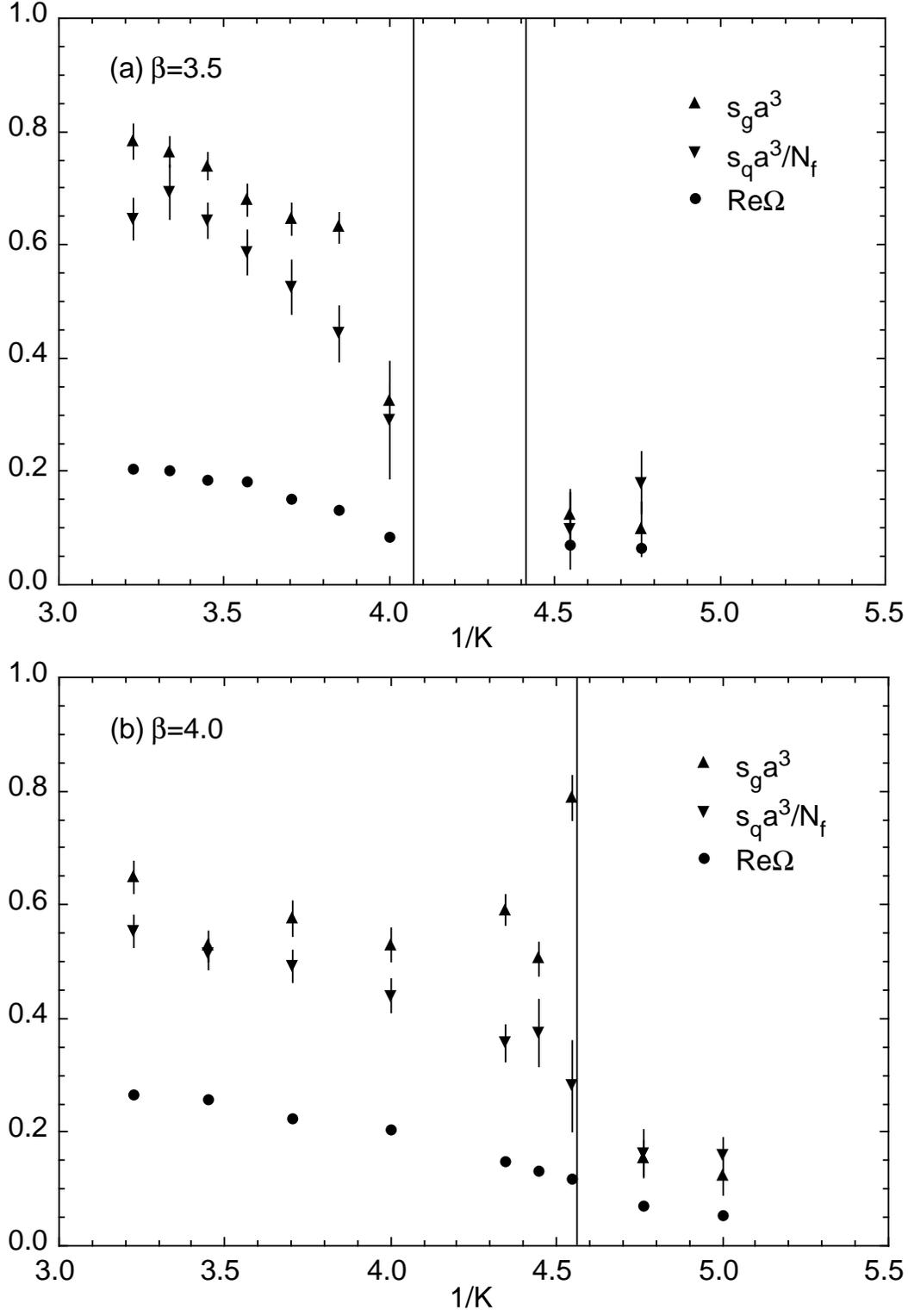

\centerline{\epsfxsize=14cm \epsfbox{fig3a.epsf}}
\centerline{\epsfxsize=14cm \epsfbox{fig3b.epsf}}
\caption{Real part of Polyakov line($Re\Omega$), gluon ($s_g$) and quark
($s_q$)
entropy density in
lattice units for $N_f=2$ full QCD with Wilson quark action obtained on an
$8^3\times 4$ lattice. }
\label{fig:fig3}\end{figure}

The validity of this consideration is confirmed through an examination of
thermodynamic observables.  In Fig.~\ref{fig:fig3}(a) we
plot the real part of the Polyakov line and quark and gluon entropy densities
in lattice units as
a function of $1/K$ at $\beta=3.5$.   The two vertical lines show the position
of
the critical line estimated from the pion  screening mass.  For small values of
$1/K$ the observables take large values typical of the high temperature phase.
However, toward the critical line at $K_c\approx 0.2454$,  they decrease,
becoming roughly similar in magnitude to those on the other side of
the conventional critical
line at $K_c\approx 0.2267$, which is in the low-temperature phase.

Results for $\beta=4.0$ are shown in Fig.~\ref{fig:fig3}(b).  The vertical line
marks the  point where the two linear extrapolation of the pion screening mass
squared in Fig.~\ref{fig:fig2}(b) cross  each other.
The increase of the three quantities across the
line shows that the thermal line comes close to the critical line at
$\beta=4.0$.

\begin{figure}[h]
\centerline{\epsfxsize=14cm \epsfbox{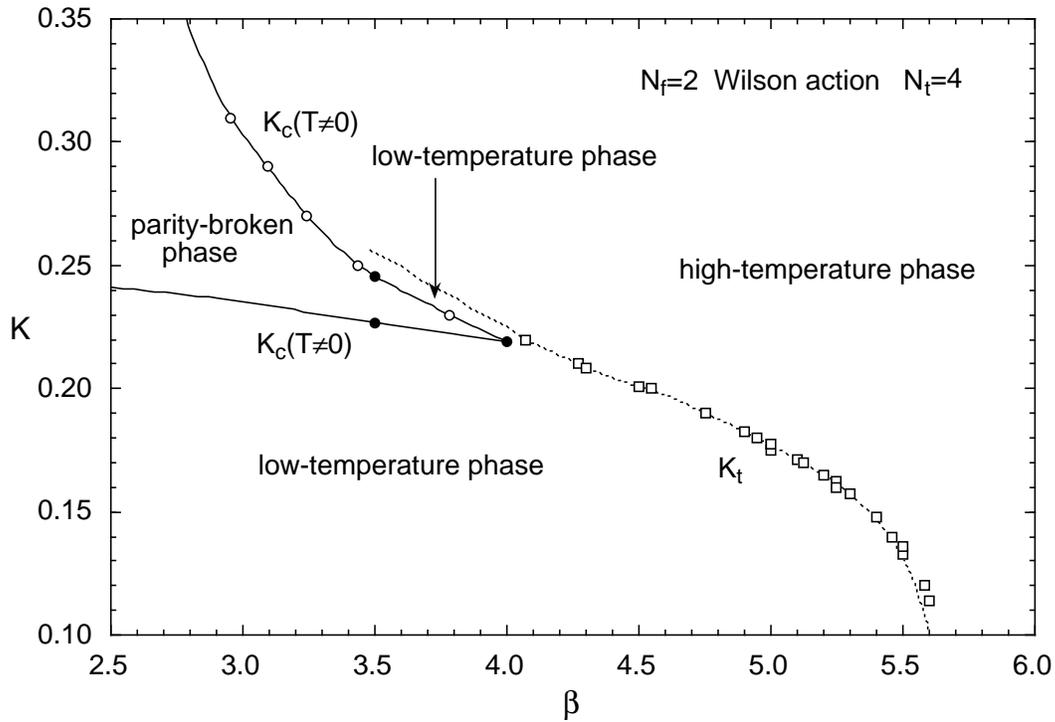}}
\caption{Phase structure of $N_f=2$ lattice QCD with Wilson quark action.
Estimates of the critical line for $N_t=4$ (solid line)
and the thermal line for chiral transition (dotted line) are shown.
For further details, see text. }
\label{fig:fig4}
\end{figure}

In Fig.~\ref{fig:fig4} we summarize our findings for the phase structure of
lattice QCD with two flavors of dynamical Wilson quarks on an $N_t=4$ lattice.
Solid circles represent
values of $K_c(\beta)$ extracted from pion screening mass.  We have tentatively
assumed that the critical line turns back at $\beta=4.0$.  Open  circles are
estimates obtained by extrapolating the inverse of the number
of conjugate gradient iterations in hybrid Monte Carlo runs to zero.  The solid
line, smoothly interpolating circles, represents one of the cusps of
the finite-temperature critical line. (We expect the tip of the cusp to be
actually
rounded. See Fig.~\ref{fig:fig1}.)  The
thermal line is indicated by the dotted line, which is an interpolation of
estimates from previous work (open squares)\cite{earlywork,qcdpaxtwo,milc},
continued toward larger values of
$K$ following our results.  As we have argued the low-temperature phase goes
around the tip of the cusp and extends into the region close to the upper
part of the critical line.

We emphasize that the zero-temperature critical line is absent from the point
of
view of finite-temperature partition function.  For
this reason we have not drawn the line in Fig.~\ref{fig:fig4}.
We note, however, that available
results indicate that this line practically forms a smooth continuation of the
lower part of the finite-temperature critical line toward weak coupling.

Let us comment here on the report of the QCDPAX Collaboration that
the thermal line crosses the critical line at
$\beta\approx 3.9-4.0$\cite{qcdpaxtwo}.  In our terminology
their result is based on simulations carried out along
the {\it zero-temperature} critical line from weak toward strong coupling.
In view of the smooth connection of the zero- and finite-temperature critical
lines
noted above, the thermal line has to cross the zero-temperature critical line,
which is the phenomenon suggested by the QCDPAX Collaboration.
In this sense their result is consistent with our phase diagram.
However, we emphasize that this crossing point does not correspond to a
singularity
of the partition function for a finite temporal lattice size.

So far we have examined the case of $N_t=4$.  For larger values of
$N_t$, we expect the thermal line and the cusp of the critical
line to move toward larger $\beta$ with the latter closely following the
zero-temperature critical line.
The distance between the thermal line and the tip of the cusp
will diminish probably as $O(a)\approx O(1/N_t)$.  An important point to note
is that
the presence of explicit chiral symmetry breaking and simulation results
including ours strongly indicate that the thermal line for the $N_f=2$ case is
a
continuous crossover and not a true phase transition\cite{comment}.
Thus, a second-order
chiral phase transition, as suggested by continuum sigma model analyses for
this case\cite{wilczek}, would emerge only in the continuum limit
$N_t\to\infty$.

Another interesting question is how our results are modified for the case of
$N_f\geq 3$. Previous simulations support a first-order chiral phase
transition\cite{qcdpaxtwo} in agreement with sigma model analyses.  This
implies
that the thermal line turns into a line of first-order phase transition
near the tip of the cusp.  The line will not cross the critical line for a
finite temporal lattice size since our reasoning for the $N_f=2$
case also applies for $N_f\geq 3$.

Let us finally add that we expect the qualitative features of the phase
structure
we found to hold for a wider class of Wilson-type actions including the clover
action\cite{clover}.

\section*{Acknowledgements}

We thank Y. Iwasaki, K. Kanaya and T. Yoshi\'e for showing us their unpublished
data and for useful discussions.
Numerical calculations for the present work have been
carried out at Center for Computational Physics and on VPP500/30 at Science
Information Processing Center, both at University of Tsukuba.
This work is supported in part by the Grants-in-Aid of
the Ministry of Education(Nos. 04NP0701, 06640372).

\newpage


\newpage
\vspace*{5cm}

 \vfill


\begin{thebibliography}{99}
\bibitem{comment}We consider that the first-order signal reported by the MILC
Collaboration\cite{milc} on an $N_t=6$ lattice around $\beta\approx 5.0$ is
likely
to be a lattice artifact.
\end{thebibliography}
\end{document}